\documentclass[twocolumn,showpacs,preprintnumbers,amsmath,amssymb]{revtex4}
\usepackage{graphicx}% Include figure files
\usepackage{dcolumn}% Align table columns on decimal point
\usepackage{bm}% bold math
\usepackage{color}% color
\usepackage{float}

\begin{document}

%\preprint{APS/123-QED}

\title{Investigation of complete and incomplete fusion in $^{7}$Li+$^{124}$Sn reaction around Coulomb barrier energies}% Force line breaks with \\

\author{V. V. Parkar$^{1,2}$\footnote{vparkar@barc.gov.in}}
\author{Sushil K. Sharma$^2$\footnote{Present address~:~The Marian Smoluchowski Institute of Physics, Jagiellonian University, {\L}ojasiewicza 11, 30-348 Krak\'ow, Poland}}
\author{R. Palit$^2$}
\author{S. Upadhyaya$^{3}$\footnote{Present address~:~The Marian Smoluchowski Institute of Physics, Jagiellonian University, {\L}ojasiewicza 11, 30-348 Krak\'ow, Poland}}
\author{A. Shrivastava$^{1,4}$}
\author{S. K. Pandit$^{1,4}$}
\author{K. Mahata$^{1,4}$}
\author{V. Jha$^{1,4}$}
\author{S. Santra$^{1,4}$}
\author{K. Ramachandran$^{1}$}
\author{T. N. Nag$^5$}
\author{P. K. Rath$^6$}
\author{Bhushan Kanagalekar$^7$}
\author{T. Trivedi$^8$}
\affiliation{$^1$Nuclear Physics Division, Bhabha Atomic Research Centre, Mumbai - 400085, India}
\affiliation{$^2$Department of Nuclear and Atomic Physics, Tata Institute of Fundamental Research, Mumbai - 400005, India}
\affiliation{$^3$Department of Applied Physics, Amity University, Noida -201313, India}
\affiliation{$^4$Homi Bhabha National Institute, Anushaktinagar, Mumbai - 400094, India}
\affiliation{$^5$Radiochemistry Division, Bhabha Atomic Research Centre, Mumbai - 400085, India}
\affiliation{$^6$Manipal Centre for Natural Sciences, Manipal University, Manipal - 576104, India}
\affiliation{$^7$Department of Physics, Rani Channamma University, Belagavi - 591156, India}
\affiliation{$^8$Department of Pure and Applied Physics, Guru Ghasidas Vishwavidyalaya, Bilaspur - 495009, India}

%\date{\today}% It is always \today, today,
             %  but any date may be explicitly specified

\begin{abstract}
The complete and incomplete fusion cross sections for $^{7}$Li+$^{124}$Sn reaction were measured using online and offline characteristic $\gamma$-ray detection techniques. The complete fusion (CF) cross sections at energies above the Coulomb barrier were found to be suppressed by $\sim$ 26 \% compared to the coupled channel calculations. This suppression observed in complete fusion cross sections is found to be commensurate with the measured total incomplete fusion (ICF) cross sections. There is a distinct feature observed in the ICF cross sections, i.e., $\textit{t}$-capture is found to be dominant than $\alpha$-capture at all the measured energies. A simultaneous explanation of complete, incomplete and total fusion (TF) data was also obtained from the calculations based on Continuum Discretized Coupled Channel method with short range imaginary potentials. The cross section ratios of CF/TF and ICF/TF obtained from the data as well as the calculations showed the dominance of ICF at below barrier energies and CF at above barrier energies.

\end{abstract}

% PACS, the Physics and Astronomy
                             % Classification Scheme.
%\keywords{Suggested keywords}%Use showkeys class option if keyword
                              %display desired
\maketitle

\section{\label{sec:Intro} Introduction}
The study of fusion involving weakly bound projectiles is of interest for probing the influence of low lying states in the continuum, the extended shape, and quantum tunneling at energies near the Coulomb barrier~\cite{Canto15}. In this context, the fusion reactions with radioactive ion beams is a topic of discussion  over the last two decades for its possible application in production of super-heavy nuclei. It is expected that the extended structure of the loosely bound nuclei could in principle induce a large enhancement of fusion which may aid to the synthesis of super-heavy nuclei in fusion reactions. Alternately, for the weakly bound  nuclei, the fusion process might be affected by their low binding energy, which can cause them to break up while approaching the fusion barrier. This may effectively reduce the complete fusion cross sections, making it difficult to form the super-heavy nuclei~\cite{Kee09,Kol16}.

Recent studies on fusion with weakly bound stable projectiles ($^{6,7}$Li and $^9$Be) on different targets have shown that the process of complete fusion (CF), where the entire projectile or all its fragments are captured, is suppressed when compared to predictions based on Coupled-channels model at energies above the Coulomb barrier~\cite{Canto15}. In particular, experiments with $^{6,7}$Li and $^9$Be projectiles on medium and heavy mass targets have given interesting conclusions on the systematics of CF suppression factor. The  suppression in CF involving these projectiles is found to be independent of target mass in many studies \cite{Gas09,vvp10,Wang14,Kundu16}.  Further the suppression factor shows an increasing trend, with decrease in the breakup threshold of the projectile \cite{Wang14}.

The observed suppression in CF could be attributed to processes where only a part of the projectile fuses with the target, known as incomplete fusion (ICF). In addition, ICF can also accommodate the two/three step processes, $\it{viz.}$; transfer of few nucleons to/from the projectile, which breaks and one of the two fragments get captured in the target. Influence of all such breakup processes on suppression in CF cross sections were discussed in recent works~\cite{Luong,Kalkal,Cook}. For investigating  the extent to which ICF influences the suppression in CF, a simultaneous measurements of both CF and ICF is crucial. At present such information is available for very limited cases~\cite{Dasgupta2004,Shrivastava13,Palsh14,Broda}.

In this paper, we report the measurement of complete and incomplete fusion cross sections for $^{7}$Li+$^{124}$Sn reaction around the Coulomb barrier energies, utilizing  online and offline characteristic $\gamma$-ray detection techniques. The dominant evaporation residues (ERs) from  complete fusion are $^{126-128}$I (3n-5n). In addition, we have also identified the residues from $\alpha$-capture, populating $^{126,127}$Te in the online measurement. In the present case, the residues  $^{128}$I (3n) and $^{126}$I (5n)  along with the residues following $\textit{t}$-capture, $\textit{viz.}$, $^{124}$Sn(t,1n)$^{126}$Sb, $^{124}$Sn(t,2n)$^{125}$Sb, $^{124}$Sn(t,3n)$^{124}$Sb and transfer products $^{124}$Sn($^{7}$Li,$^{6}$Li)$^{125}$Sn (one neutron stripping) and $^{124}$Sn($^{7}$Li,$^{8}$Li)$^{123}$Sn (one neutron pickup) undergo radioactive  decay with half-lives suitable for offline measurements. The offline $\gamma$-ray activity measurements were carried out at few energies for extraction of cross sections of these residues to get complete information of total ICF and transfer channels. For some nuclei, it was possible to obtain cross sections using both in-beam and off-beam methods. The statistical model and coupled channel calculations were also carried out.

The paper is organized as follows: the experimental details are described in section~\ref{sec:Expt}. The measured CF and ICF cross sections are compared with coupled channel calculations in section~\ref{sec:Result}. The summary of the present study is given in section~\ref{sec:Sum}.

\section{\label{sec:Expt} Experimental Details}
The measurements were carried out at 14UD BARC-TIFR Pelletron-Linac accelerator facility, Mumbai using $^7$Li beam. The details of online and offline $\gamma$-ray measurement methods are given here.
\subsection{Online $\gamma$-ray Measurement}
\begin{figure}
\includegraphics[width=0.72\textwidth,trim=0.6cm 8.7cm 2.5cm 0.2cm, clip=true]{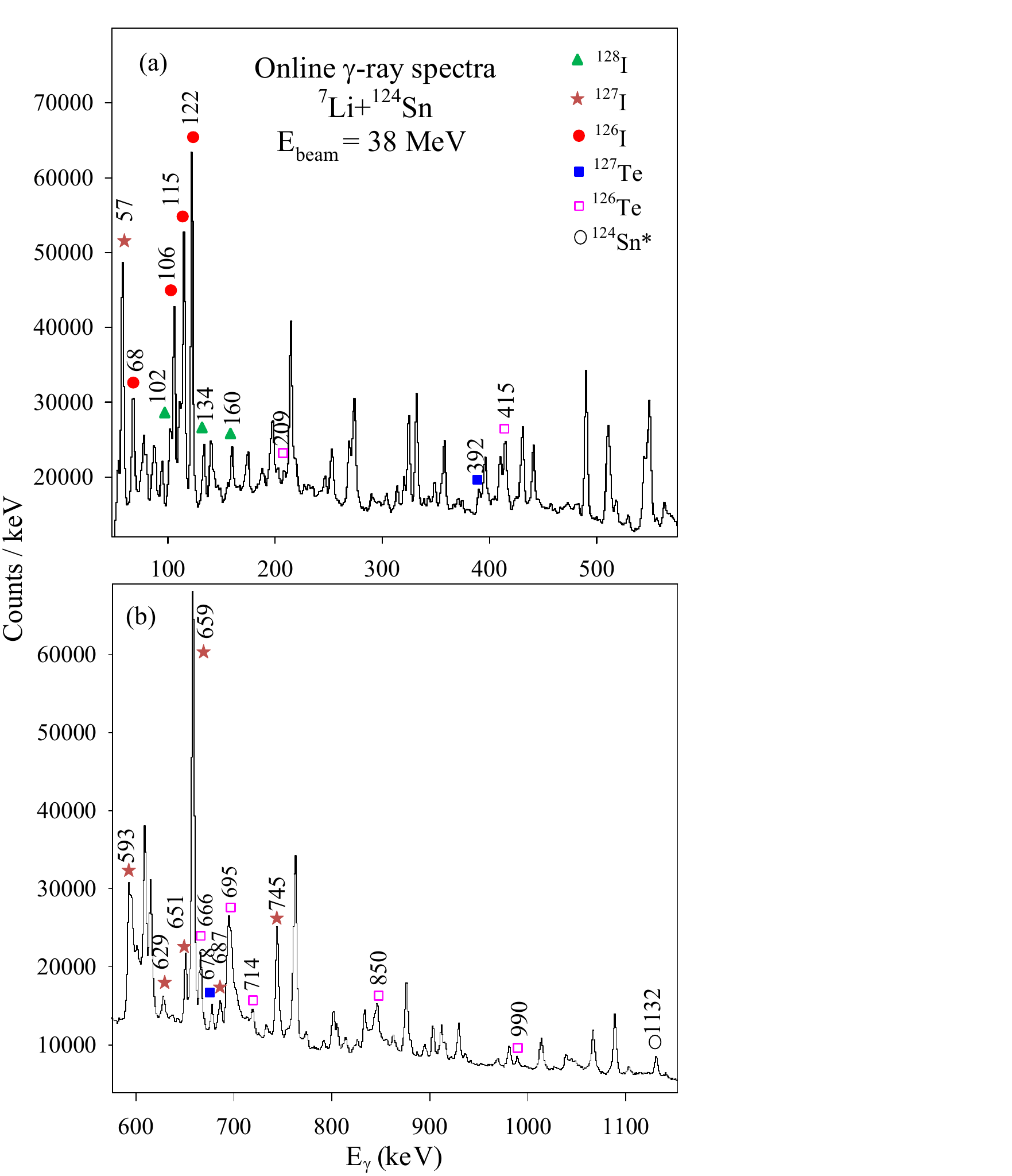}
\caption{\label{fig1} $\gamma$-ray addback spectrum from the clover detector at 125$^{\circ}$ obtained in $^7$Li+$^{124}$Sn reaction at E$_{\textrm{beam}}$ = 38 MeV. The $\gamma$ lines from the possible evaporation residues ($^{126,127,128}$I) following CF are labeled. Also the $\gamma$ lines following the $\alpha$-capture channel ($^{126,127}$Te) and inelastic $^{124}$Sn$^*$ are marked.}
\end{figure}

A detailed description of the experimental setup used for online $\gamma$-ray measurements was given in our earlier work \cite{vvp10} and only a short summary pertinent to this work is presented here. The $^7$Li beam with energies E$_{\textrm{beam}}$ = 17-39 MeV in one MeV step was bombarded on $^{124}$Sn target (thickness = 2.47 $\pm$ 0.04 mg/cm$^2$). The beam energies were corrected for the loss at half the target thickness and used in the further analysis. Two Compton suppressed clover detectors were placed at a distance of 25 cm from the target centre, one at 125$^{\circ}$, for the estimation of absolute cross section of populated reaction channels and other at 90$^{\circ}$, for identification of unshifted $\gamma$ lines. The absolute efficiency of both the detectors was determined using a set of radioactive $^{152}$Eu, $^{133}$Ba and $^{241}$Am sources mounted in the same geometry as the target. Along with the clover detectors, one monitor detector (= 500 $\mu$m) was placed at 30$^{\circ}$. The monitor detector was utilised in the ER cross section estimation using the measured elastic (Rutherford) scattering cross section. The integrated beam current deposited at the beam dump after the target was also recorded using the high precision current integrator. Figure~\ref{fig1} shows the typical $\gamma$-ray addback spectrum from the clover kept at 125$^{\circ}$ and E$_{\textrm{beam}}$ = 38 MeV for $^{7}$Li+$^{124}$Sn reaction. The $\gamma$ lines from the possible ERs following CF; $\textit{viz}$, $^{126-128}$I are labeled. Also the identified $\gamma$ lines following the ICF channel; $\textit{viz}$, from $\alpha$-capture, $^{126,127}$Te are marked. The $\textit{t}$-capture process populates $^{124-126}$Sb nuclei, of which $^{124}$Sb and $^{126}$Sb have metastable states of few minutes. Furthermore $^{125}$Sb level structure is not well studied in literature. Hence, it is difficult to measure their cross sections accurately in the online measurements.

\subsection{Offline $\gamma$-ray Measurement}
\begin{figure}
\includegraphics[width=0.56\textwidth,trim=0.71cm 15.9cm 3.4cm 1cm, clip=true]{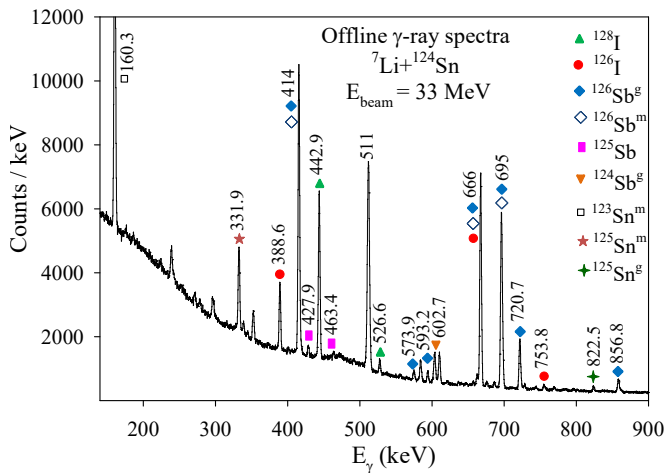}
\caption{\label{fig2} Offline $\gamma$-ray spectrum obtained in HPGe detector for $^7$Li+$^{124}$Sn reaction at E$_{\textrm{beam}}$ = 33 MeV. Identified $\gamma$ lines from different residues following CF ($^{126,128}$I), $\textit{t}$-capture ($^{124,125,126}$Sb), one neutron stripping ($^{125}$Sn) and one neutron pickup ($^{123}$Sn) are marked.}
\end{figure}

Six targets of $^{124}$Sn having thicknesses in the range of 1.5-4.0 mg/cm$^2$ were irradiated with beam of $^7$Li at 19.3, 22.3, 24.8, 28.8, 33 and 35.9 MeV energies. These energies were chosen in such a way that after energy loss correction at half the target thickness, they match with that of previously measured online $\gamma$-ray measurements. The targets with the Al catcher ($\sim$ 1 mg/cm$^2$ thick) were placed normal to the beam direction so that the recoiling residues are stopped in target-catcher assembly. The irradiation time was typically 7-18 hrs from highest to lowest bombarding energy. The beam current was $\sim$ 10-80 nA. To monitor current variations during each irradiation, a CAMAC scaler was utilized which recorded the integrated current in an intervals of 1 min. The irradiated target-catcher assembly was then sticked to the perspex sheet and the sheet was kept at a fixed distance ($\sim$ 10 cm) in front of the HPGe detector. The HPGe detector was surrounded by 2 mm thick Cu and Cd sheets and 5 cm thick Pb sheets to reduce the background. The energy calibration and absolute efficiency of the HPGe detector was measured by using a set of calibrated radioactive $^{152}$Eu, $^{133}$Ba and $^{241}$Am sources placed at the same geometry as the target. All the six targets were counted individually at various intervals following the half lives. The residues from CF, ICF and transfer reactions were identified by the characteristic $\gamma$ lines emitted by their daughter nuclei as shown in Fig.~\ref{fig2} and listed in Table I.

\begin{table}
\caption{\label{tab:table1}List of identified residues in the offline $\gamma$-ray measurement for the $^7$Li+$^{124}$Sn reaction along with their radioactive decay half-lives (T$_{1/2}$), $\gamma$-ray energies and intensities following their decays \cite{Nudat}.}
\begin{tabular}{ccccc}
\hline
Reaction & ER & T$_{1/2}$ & E$_\gamma$ (keV) & I$_\gamma$ (\%) \\ \hline \\
$^{124}$Sn($^7$Li,3n) & $^{128}$I & 24.99 min & 442.9 & 12.6 \\
& & & 526.6 & 1.2 \\
$^{124}$Sn($^7$Li,5n) & $^{126}$I & 12.93 d & 388.6 & 35.6 \\
& & & 753.8 & 4.2 \\
$^{124}$Sn(t,1n) & $^{126}$Sb$^g$ & 12.35 d & 414.7 & 83.3 \\
& & & 573.9 & 6.7 \\
& & & 593.2 & 7.5 \\
& & & 666.5 & 99.6 \\
& & & 695.0 & 99.6 \\
& & & 697.0 & 29.0 \\
& & & 720.7 & 53.8 \\
& & & 856.8 & 17.6 \\
& $^{126}$Sb$^m$ & 19.15 min & 414.5 & 86.0 \\
& & & 666.1 & 86.0 \\
& & & 694.8 & 82.0 \\
& & & 928.2 & 1.3 \\
& & & 1034.9 & 1.8 \\
$^{124}$Sn(t,2n) & $^{125}$Sb & 2.76 yr & 427.9 & 29.6 \\
& & & 463.4 & 10.5 \\
& & & 600.6 & 17.7 \\
& & & 636.0 & 11.2 \\
$^{124}$Sn(t,3n) & $^{124}$Sb$^g$ & 60.20 d & 602.7 & 97.8 \\
& & & 1691.0 & 47.6 \\
$^{124}$Sn($^7$Li,$^6$Li) & $^{125}$Sn$^g$ & 9.64 d & 822.5 & 4.3 \\
& & & 915.6 & 4.1 \\
& & & 1067.1 & 10.0 \\
& & & 1089.2 & 4.6 \\
& $^{125}$Sn$^m$ & 9.52 min & 331.9 & 97.3 \\
$^{124}$Sn($^7$Li,$^8$Li) & $^{123}$Sn$^m$ & 40.06 min & 160.3 & 85.7 \\ \hline
\end{tabular}
\end{table}

\section{\label{sec:Result} Results and Discussion}
\subsection{Data Reduction}
\subsubsection{\label{online} Online $\gamma$-ray Analysis}
The emission cross sections for $\gamma$ transitions of interest for online measurements were calculated from the relation
\begin{equation}
\sigma_\gamma = \frac{{Y_\gamma  }}{{Y_M }} \frac{{d\Omega _M }}{{\epsilon_\gamma  }} \frac{d\sigma_{Ruth}}{d\Omega}
\end{equation}

where $Y_\gamma$ is the $\gamma$-ray yield after correcting for the internal conversion, $Y_M$ is the monitor yield, $d\Omega _M$ is the solid angle of the monitor detector, $\epsilon_\gamma$ is the absolute efficiency of the detector for a particular $\gamma$-ray energy, and $\frac{d\sigma_{Ruth}}{d\Omega}$ is the Rutherford cross section (at $\theta_M$ = 30$^\circ$) at the same beam energy. For $^{126-128}$I and $^{127}$Te nuclei, all the cross sections of $\gamma$ transitions feeding to the ground and metastable (having $\sim$ few $\mu$s life times) states of the particular residue are added to get the residue cross sections. The $\gamma$ lines populating the ground and metastable states in these nuclei are taken from Refs.\ \cite{126I,127I,128I,126Te}. In the case of even-even $^{126}$Te nucleus, the identified $\gamma$ lines \cite{126Te} also have the contribution from offline decay events of $^{126}$Sb$^m$ (t$_{1/2}$ = 19.15 min) which were formed after triton capture followed by one neutron evaporation. Hence to extract the cross section of $^{126}$Te, we have estimated the contribution from $^{126}$Sb$^m$ decay, for which the cross section was measured from offline counting at few energies (explained in the next section) and interpolated for the intermediate energies and the corrected yield for the particular $\gamma$ transition was used. Here, only the ground state transition 2$^+$ $\rightarrow$ O$^+$ (666 keV) is used to get the $^{126}$Te cross section.

\begin{figure}
\includegraphics[width=0.48\textwidth,trim=4cm 12.2cm 3.0cm 6cm, clip=true]{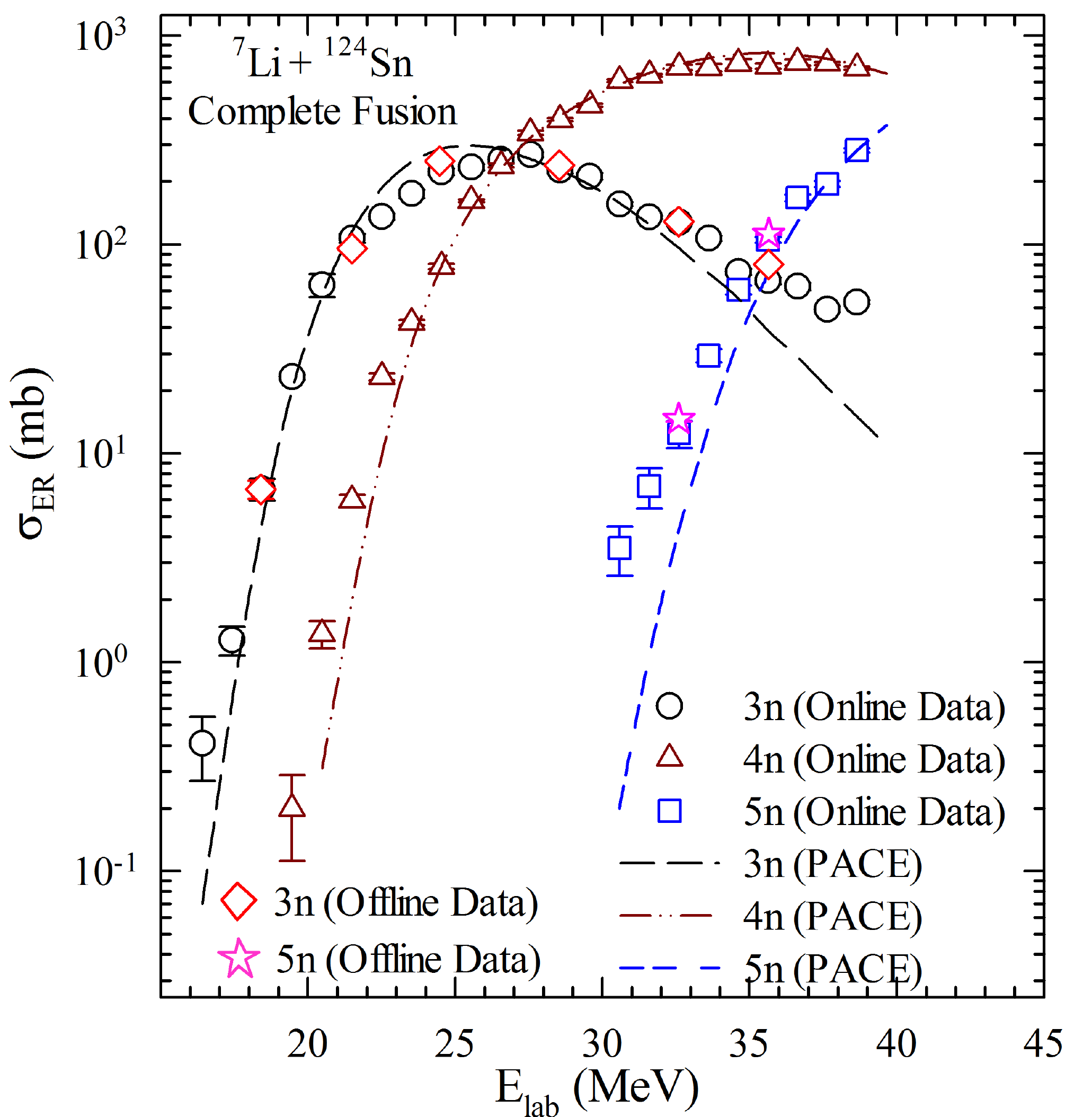}
\caption{\label{7Li_ER} ER cross sections from online $\gamma$-ray measurement for 3n ($^{128}$I), 4n ($^{127}$I), and 5n ($^{126}$I) channels following CF are represented by open circles, open triangles, and open squares, respectively. The ER data from offline $\gamma$ ray measurement for 3n and 5n channels are shown by open diamonds and open stars respectively. The results of the statistical model calculations for the corresponding ERs are shown by long dashed (3n), dashed dot dot (4n), and short dashed (5n) lines.}
\end{figure}

The cross sections for $^{128}$I (3n), $^{127}$I (4n), and $^{126}$I (5n) ERs following CF for $^7$Li+$^{124}$Sn reaction are shown by open circle, open triangle and open square symbols respectively along with the statistical model predictions using {\sc PACE} code \cite{Gav80} in Fig.~\ref{7Li_ER}. The error bars on the data are due to statistical errors in the determination of the $\gamma$-ray yields, background subtraction and absolute efficiency of the detectors. In the {\sc PACE} calculations, the cross section for each partial wave ($\textit{l}$ distribution) obtained from the Coupled Channel (CC) calculation code {\sc CCFULL} \cite{Hag99} were fed as an input. The default optical potentials available in the code were used. The only free parameter remaining in the {\sc PACE} input was the level density parameter `a', which showed a negligible dependence on the values between a = A/9 and a = A/10. The complete fusion cross sections were determined by dividing the cumulative measured ($\sigma_{3n+4n+5n}^{expt}$) cross sections by the ratio R, which gives the missing ER contribution, if any. Here the ratio R is defined as \begin{math}{\rm{R = }}\sum\limits_{\rm{x}} {{\rm{\sigma }}_{_{{\rm{xn}}} }^{{\rm{PACE}}} } {\rm{/\sigma }}_{_{{\rm{fus}}} }^{{\rm{PACE}}}\end{math}, where x = 3, 4, 5. The ratio (R) and the CF cross sections thus obtained are listed in Table II.

\begin{table}[htbp]
\caption{Measured cross sections for $\Sigma\sigma_{xn}$(x = 3, 4, 5) evaporation residues and complete fusion along with the ratio R, obtained from PACE (defined in the text) for $^7$Li+$^{124}$Sn reaction for the measured energy range.}
\begin{tabular}
{ccccc}
\hline
E$_{lab}$ & E$_{c.m.}$ & $\sigma_{3n+4n+5n}^{expt}$ & R(PACE) & $\sigma_{CF}^{expt}$ \\
(MeV) & (MeV) & (mb) & & (mb) \\
\hline
16.4 & 15.5 & 0.41 $\pm$ 0.14 & 0.64 & 0.64 $\pm$ 0.22 \\
17.4 & 16.5 & 1.28 $\pm$ 0.20 & 0.76 & 1.69 $\pm$ 0.26 \\
18.4 & 17.5 & 6.76 $\pm$ 0.81 & 0.89 & 7.60 $\pm$ 0.91 \\
19.5 & 18.4 & 23.4 $\pm$ 0.6 & 0.93 & 25.1 $\pm$ 0.7 \\
20.5 & 19.4 & 65.4 $\pm$ 8.0 & 0.96 & 68.3 $\pm$ 8.4 \\
21.5 & 20.3 & 113 $\pm$ 5 & 0.97 & 116 $\pm$ 5 \\
22.5 & 21.3 & 159 $\pm$ 5 & 0.98 & 162 $\pm$ 5 \\
23.5 & 22.3 & 217 $\pm$ 8 & 0.99 & 220 $\pm$ 8 \\
24.5 & 23.2 & 302 $\pm$ 8 & 0.99 & 305 $\pm$ 8 \\
25.5 & 24.2 & 396 $\pm$ 8 & 0.99 & 400 $\pm$ 9 \\
26.6 & 25.1 & 493 $\pm$ 10 & 0.99 & 498 $\pm$ 10 \\
27.6 & 26.1 & 611 $\pm$ 12 & 0.99 & 617 $\pm$ 12 \\
28.6 & 27.0 & 622 $\pm$ 11 & 0.99 & 628 $\pm$ 11 \\
29.6 & 28.0 & 676 $\pm$ 12 & 0.99 & 683 $\pm$ 12 \\
30.6 & 29.0 & 768 $\pm$ 10 & 0.99 & 776 $\pm$ 10 \\
31.6 & 29.9 & 787 $\pm$ 11 & 0.99 & 796 $\pm$ 11 \\
32.6 & 30.9 & 845 $\pm$ 21 & 0.99 & 856 $\pm$ 22 \\
33.6 & 31.8 & 838 $\pm$ 29 & 0.99 & 850 $\pm$ 30 \\
34.6 & 32.8 & 868 $\pm$ 39 & 0.99 & 881 $\pm$ 39 \\
35.6 & 33.7 & 885 $\pm$ 21 & 0.98 & 900 $\pm$ 21 \\
36.6 & 34.7 & 974 $\pm$ 26 & 0.98 & 991 $\pm$ 26 \\
37.6 & 35.6 & 978 $\pm$ 15 & 0.98 & 996 $\pm$ 16 \\
38.7 & 36.6 & 1034 $\pm$ 14 & 0.98 & 1056 $\pm$ 14 \\
\hline
\end{tabular}
\label{tab:fus}
\end{table}

\subsubsection{\label{offline} Offline $\gamma$-ray Analysis}
\begin{figure}
\includegraphics[width=0.80\textwidth,trim=1.5cm 19.6cm 2.9cm 0.9cm, clip=true]{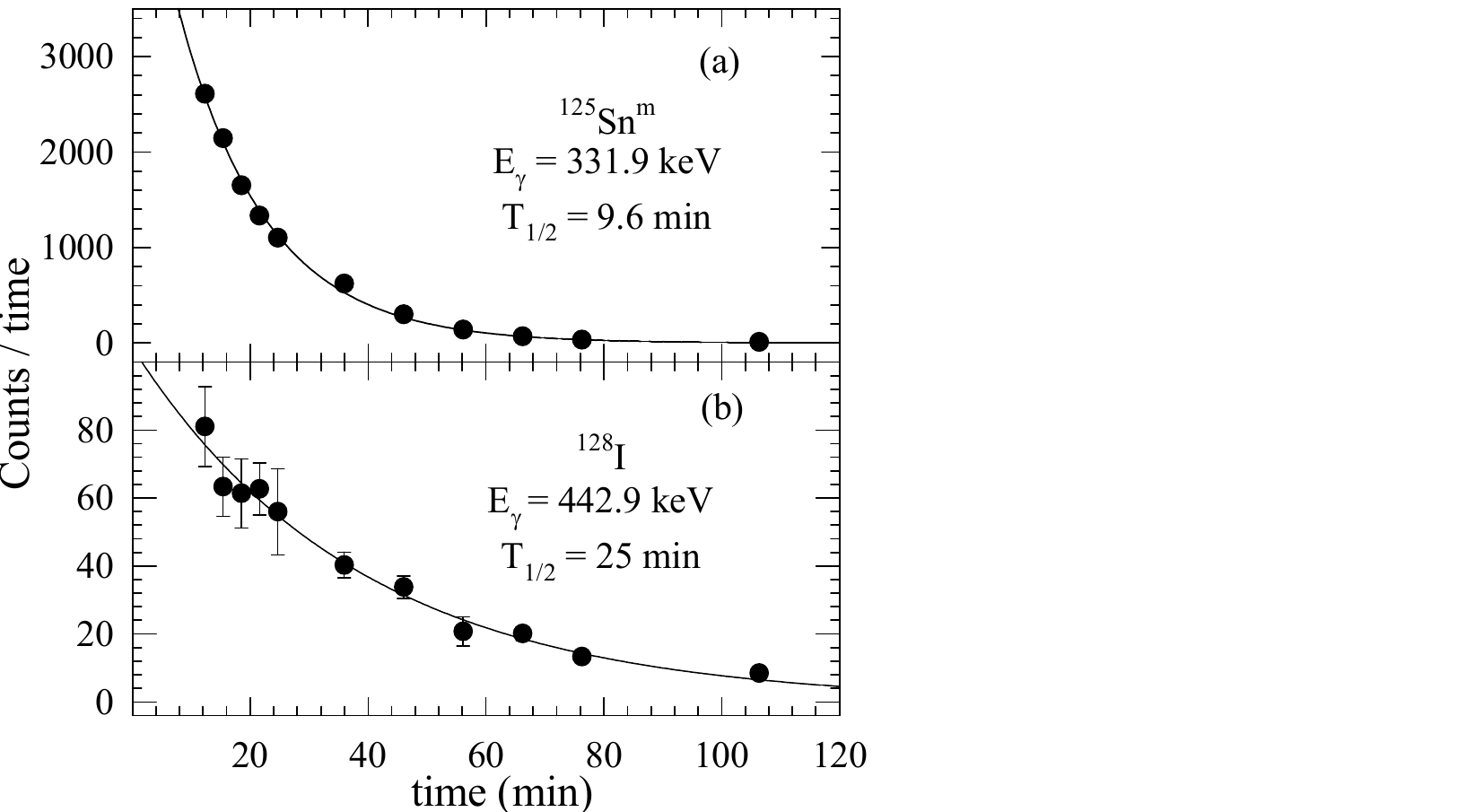}
\caption{\label{halflife} Typical radioactive decay curves obtained for (a) $^{125}$Sn$^m$ and (b) $^{128}$I residues.}
\end{figure}

For the offline $\gamma$ counting experiment, the residue cross section ($\sigma_{R}$) at a particular beam energy was obtained using the expression as follows:
\begin{equation} \label{eq2}
\sigma_{R}=\frac{Y_{\gamma}{\lambda}} {N_t{\epsilon_{\gamma}}I_{\gamma}k},
\end{equation}
\textrm{where}
\[k=\sum_{n=1}^mi_n(1-e^{{\lambda}t_{step}})(e^{-{\lambda}[t_1+(n-1)t_{step}]}-e^{-{\lambda}[t_2+(n-1)t_{step}]}),\]
here, Y$_\gamma$ is the area under the $\gamma$-peak corresponding to the residual nucleus with decay constant $\lambda$, N$_t$ is the number of target nuclei per unit area, $\epsilon_\gamma$ is the efficiency of the HPGe detector at the peak energy and I$_\gamma$ is the intensity branching ratio associated with the particular $\gamma$ line corresponding to the residual nucleus. $t_1$ and $t_2$ are the start and stop times of counting for the irradiated samples w.r.t. the beam stop, $t_{step}$ is the step size in which the current was recorded in the scaler, $i_n$ is the current recorded by the scaler at the n$^{th}$ interval, and m is the total number of intervals of irradiation. The half-lives of all the residues of our interest were confirmed by following their activities as a function of time. Typical radioactive decay curves obtained for $^{125}$Sn$^m$ and $^{128}$I residues are shown in Fig.~\ref{halflife}(a) and (b), respectively. Various $\gamma$ lines corresponding to the same residue having different I$_\gamma$ were also used for confirmation of the estimated channel cross section.

The cross sections for $^{128}$I (3n) and $^{126}$I (5n) ERs from the offline measurement are shown by open diamond and open star symbols respectively in Fig.~\ref{7Li_ER}. As can be seen from the figure, the cross sections for these two channels from offline and online $\gamma$-ray measurements showed good agreement, thus leaving no doubt about missing any major $\gamma$ line feeding the ground state in online $\gamma$ measurement. The extracted cross sections for $\textit{t}$-capture and 1n transfer channels are discussed in section~\ref{sec:ICF}.

In the offline $\gamma$-ray measurements, special care was taken to reduce the systematic uncertainties that could arise from different sources such as (i) beam current, (ii) target thickness, (iii) detector efficiency, and (iv) extraction of $\gamma$-ray yield. The current integrator was calibrated using a precision Keithley current source. Also the beam current fluctuation was recorded by dividing the irradiation time in small intervals (1 min) and was used in the analysis. This procedure reduces the uncertainty in the current measurement to less than 1\%. The target thicknesses were measured using the Rutherford backscattering method with $^{16}$O beam as well as $\alpha$ energy loss technique with Am-Pu $\alpha$ source. Uncertainty in the thickness ($\sim$ 2\%) was taken into account in the analysis for each target. The absolute detector efficiency was also measured repeatedly and found to remain invariant with time during the whole experiment. The uncertainty ($\sim$ 1\%) in the fitting parameters of the efficiency curve was taken into account. The total uncertainty on the residue cross sections were obtained after adding the statistical and the systematic errors as listed in Table~\ref{tab:table3}.

\begin{figure}
\includegraphics[width=0.54\textwidth,trim=2.9cm 13.7cm 2.0cm 5.4cm, clip=true]{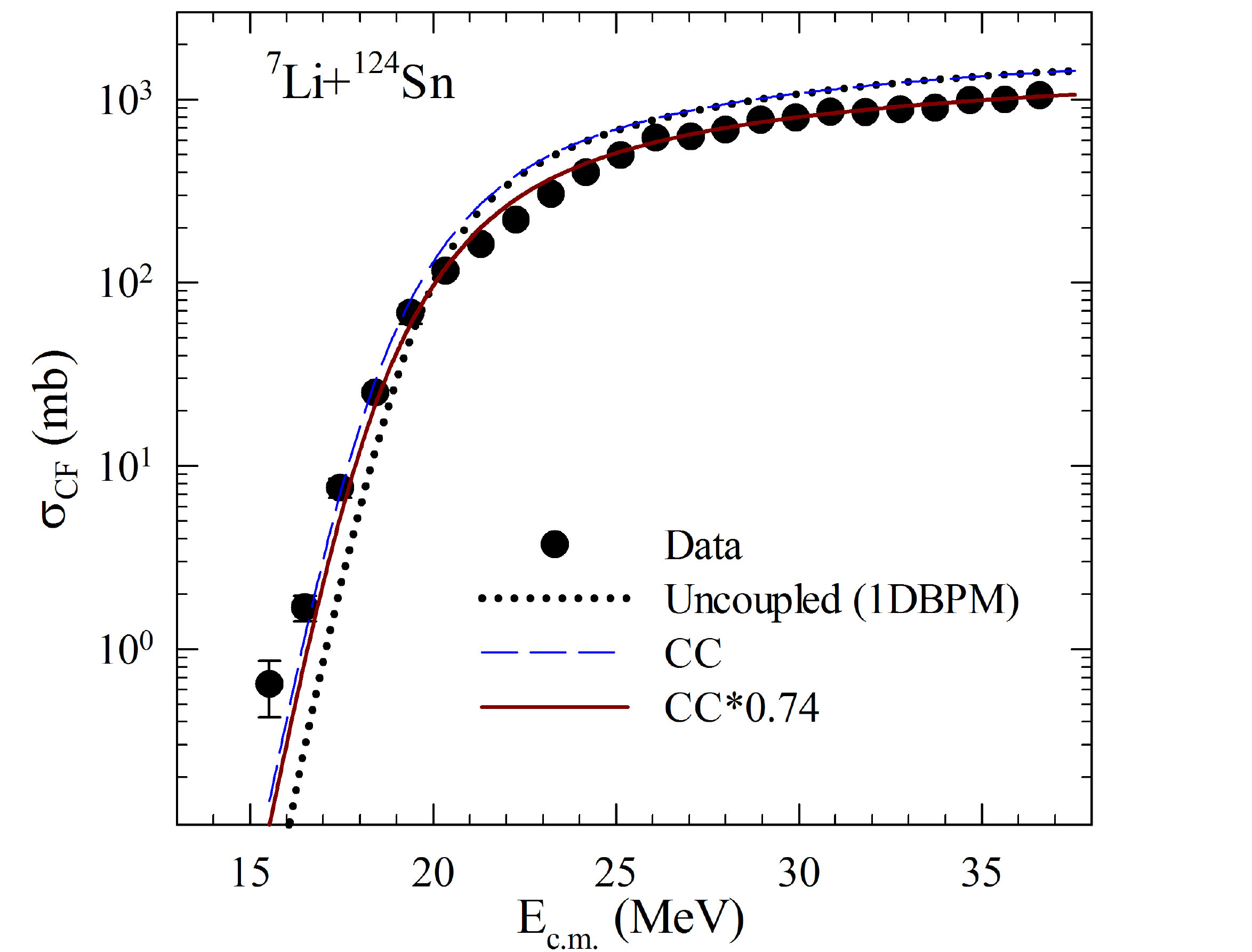}
\caption{\label{7Li_CF} Complete fusion cross section (filled circles) for the $^7$Li+$^{124}$Sn reaction compared with coupled (dashed lines) and uncoupled (dotted lines) results from CCFULL calculations. Solid lines were obtained by multiplying the coupled results by a factor of 0.74.}
\end{figure}

\subsection{\label{sec:Coupled} Coupled Channel Calculations}
Coupled channel calculations were performed using the modified version of CCFULL \cite{Hag99}, which can include the effect of projectile ground-state spin and the projectile excitation in addition to the target excitation. The potential parameters used were:
V$_0$ = 45 MeV, r$_0$ = 1.17 fm and a$_0$ = 0.62 fm, obtained from the Woods-Saxon parametrization of the Akyuz-Winther (AW) potential \cite{Bro81b}. The corresponding uncoupled barrier height V$_B$, radius R$_B$, and curvature $\hbar\omega$ derived for the present systems are 19.7 MeV, 10.3 fm and 4.13 MeV respectively. The full couplings include the coupling of the projectile ground state (3/2$^-$) and first excited state (1/2$^-$, 0.478 MeV) with $\beta_{00}$ ($\beta_2$ for the ground-state reorientation) = 1.189, $\beta_{01}$ ($\beta_2$ for the transition between the ground and the first excited states) = 1.24. These values are taken from Ref.\ \cite{Rath7Li}. Target coupling included the 3$^{-}$ vibrational excited state in $^{124}$Sn with E$_x$ = 2.603 MeV, $\beta_3$ = 0.106 \cite{kibedi}. The effect of coupling of 2$^{+}$ excited state ($\beta_2$ = 0.0953, E$_x$ = 1.132 MeV) in $^{124}$Sn is found to be less important compared to 3$^-$ state. The breakup or transfer coupling channel cannot be included in these calculations.

The results from the uncoupled and coupled calculations are shown in Fig.~\ref{7Li_CF} by dotted and dashed lines, respectively. It was observed that at sub-barrier energies, the calculated fusion cross sections with the couplings (dashed lines) are enhanced compared to the uncoupled values. However, at above-barrier energies, the calculated values of fusion with or without couplings are higher than the measured ones. It was interesting to observe that when the calculated fusion cross sections obtained with the above coupling are normalized by a factor of 0.74, the reduced fusion values (denoted by solid line) reproduce the experimental fusion cross sections very well specially at energies above the Coulomb barrier.  Thus, one can conclude that the CF cross sections in this region are suppressed by 26 $\pm$ 4\% compared to the prediction of CCFULL calculations. The uncertainty of 4\% in suppression factor was estimated from the uncertainties in V$_B$ and $\sigma_{CF}$. In recent studies \cite{Wang14,Kundu16}, the complete fusion cross section data available with weakly bound and strongly bound projectiles on various targets was shown to be systematically target independent. Present work also support this observation with $^7$Li projectile in medium and heavy mass region.

\subsection{\label{sec:ICF} ICF and 1n transfer cross sections} \label{ICF}
\begin{figure}
\includegraphics[width=0.50\textwidth,trim=2.3cm 9.7cm 2.6cm 2.5cm, clip=true]{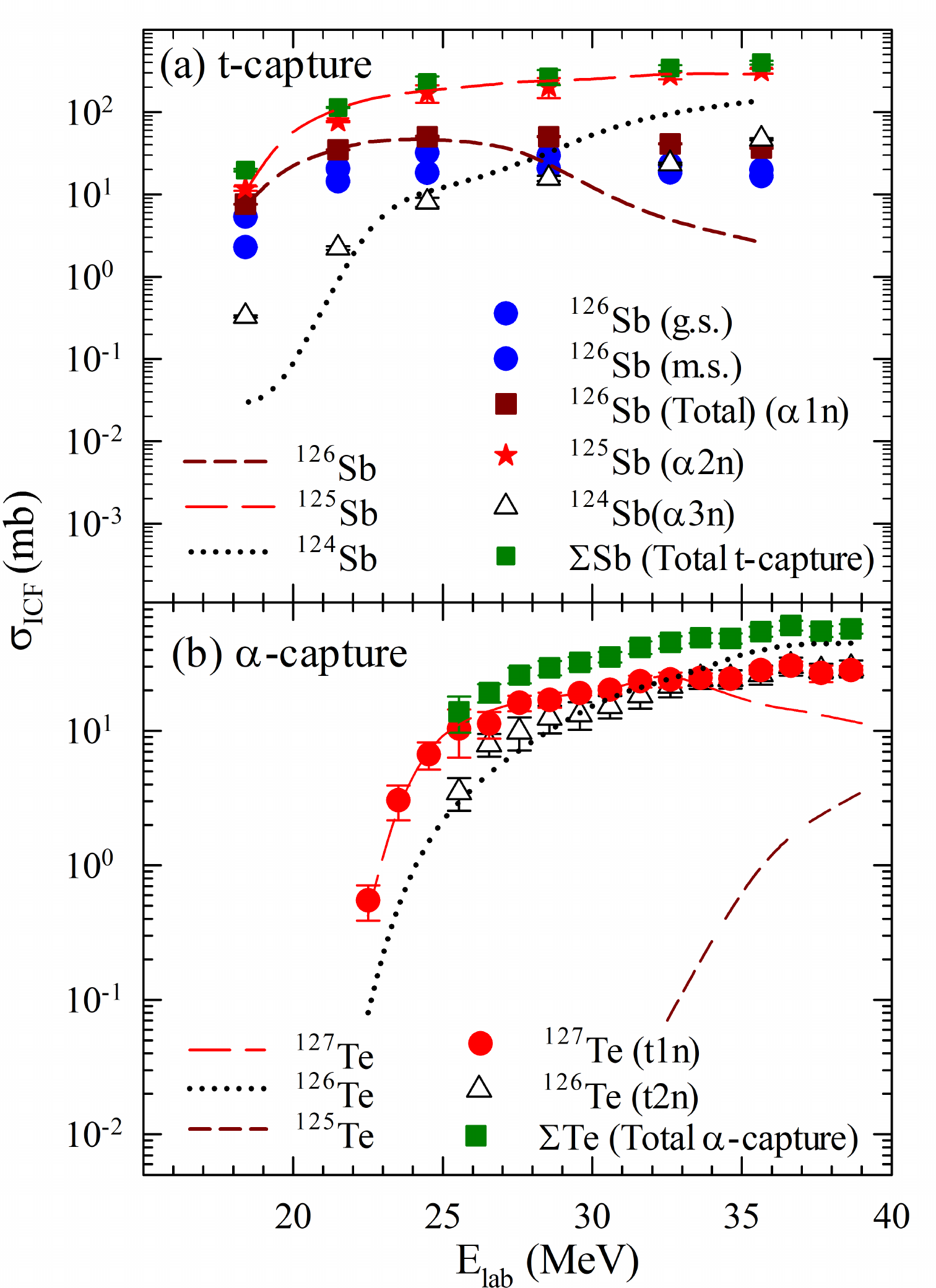}
\caption{\label{7Li_ICFdata} Measured residue cross sections for (a) $\textit{t}$-capture process (from offline $\gamma$-ray measurement) (b) $\alpha$-capture process (from online $\gamma$-ray measurement) in $^7$Li+$^{124}$Sn reaction are plotted. The lines are the predictions from statistical model calculations for the corresponding residues (see text for details).}
\end{figure}

\begin{table*}
\caption{\label{tab:table3} Measured cross sections for incomplete fusion products along with 1n pickup and 1n stripping cross sections obtained from online and offline $\gamma$-ray measurement techniques in $^7$Li+$^{124}$Sn reaction.}
\small
\begin{tabular}
{cccccccccc}
\hline \hline \\
E$_{lab}$ & $^{127}$Te & $^{126}$Te & $^{126}$Sb$^m$ & $^{126}$Sb$^g$ & $^{125}$Sb & $^{124}$Sb$^g$ & $^{125}$Sn$^g$ & $^{125}$Sn$^m$ & $^{123}$Sn$^m$ \\
(MeV) & (mb) & (mb) & (mb) & (mb) & (mb) & (mb) & (mb) & (mb) & (mb)\\
\hline \\
18.4 &  &  & 5.34 $\pm$ 0.5 & 2.28 $\pm$ 0.2 & 11.8 $\pm$ 1.7 & 0.33 $\pm$ 0.03 & 0.23 $\pm$ 0.02 & 1.18 $\pm$ 0.16 & 0.17 $\pm$ 0.01 \\
21.5 &  &  & 20.5 $\pm$ 1.0 & 14.4 $\pm$ 1.2 & 77.0 $\pm$ 3.4  & 2.24 $\pm$ 0.11 & 2.13 $\pm$ 0.18 & 4.95 $\pm$ 0.05 & 1.11 $\pm$ 0.20 \\
22.5 & 0.55 $\pm$ 0.16 &  &  &  &  &  &  &  &  \\
23.5 & 3.05 $\pm$ 0.88 & &  &  &  &  &  &  &  \\
24.5 & 6.68 $\pm$ 1.53 & & 32.1 $\pm$ 1.9 & 18.3 $\pm$ 1.5 & 172 $\pm$ 41 & 8.10 $\pm$ 1.02 & 8.54 $\pm$ 2.62 & 12.0 $\pm$ 0.9 & 5.56 $\pm$ 0.34 \\
25.5 & 10.4 $\pm$ 4.0 & 3.50 $\pm$ 0.96 & 31.5 $\pm$ 1.1$^a$  &  &  &  &  &  &  \\
26.6 & 11.3 $\pm$ 2.5 & 7.96 $\pm$ 1.53 & 31.0 $\pm$ 1.1$^a$ &  &  &  &  &  &   \\
27.6 & 16.1 $\pm$ 2.2 & 9.87 $\pm$ 2.73 & 30.5 $\pm$ 1.1$^a$ &  &  &  &  &  &   \\
28.6 & 17.0 $\pm$ 2.2 & 12.5 $\pm$ 2.9 & 29.7 $\pm$ 1.6 & 20.6 $\pm$ 1.6 & 203 $\pm$ 56 & 15.7 $\pm$ 1.1 & 13.1 $\pm$ 2.6 & 17.3 $\pm$ 1.7 & 6.42 $\pm$ 1.12 \\
29.6 & 19.0 $\pm$ 0.7 & 13.3 $\pm$ 3.1 & 28.0 $\pm$ 1.7$^a$ &  &  &  &  &  &   \\
30.6 & 20.2 $\pm$ 1.4 & 15.2 $\pm$ 2.9 & 26.3 $\pm$ 1.7$^a$ &  &  &  &  &  &   \\
31.6 & 23.3 $\pm$ 2.4 & 18.4 $\pm$ 3.8 & 24.6 $\pm$ 1.7$^a$ &  &  &  &  &  &   \\
32.6 & 24.1 $\pm$ 3.1 & 21.6 $\pm$ 3.8 & 22.9 $\pm$ 1.3 & 18.4 $\pm$ 1.4 & 279 $\pm$ 29 & 23.6 $\pm$ 1.8 & 22.5 $\pm$ 1.3 & 20.8 $\pm$ 1.3 & 8.18 $\pm$ 1.05 \\
33.6 & 24.7 $\pm$ 2.1 & 24.5 $\pm$ 3.9 & 21.9 $\pm$ 1.6$^a$ &  &  &  &  &  &   \\
34.6 & 24.3 $\pm$ 1.6 & 24.4 $\pm$ 3.8 & 20.9 $\pm$ 1.6$^a$ &  &  &  &  &  &   \\
35.6 & 28.3 $\pm$ 1.9 & 26.3 $\pm$ 4.3 & 19.9 $\pm$ 1.5 & 16.7 $\pm$ 1.5 & 316 $\pm$ 22 & 46.9 $\pm$ 2.6 & 25.2 $\pm$ 1.6 & 24.6 $\pm$ 1.8 & 17.0 $\pm$ 1.2 \\
36.6 & 30.5 $\pm$ 2.3 & 30.4 $\pm$ 4.6 & 18.9 $\pm$ 1.5$^b$ &  &  &  &  &  &   \\
37.6 & 26.9 $\pm$ 3.8 & 28.0 $\pm$ 3.5 & 17.9 $\pm$ 1.5$^b$ &  &  &  &  &  &   \\
38.7 & 28.3 $\pm$ 2.1 & 29.2 $\pm$ 4.2 & 16.9 $\pm$ 1.5$^b$ &  &  &  &  &  &   \\
\hline \hline
\end{tabular}
a~:~interpolated value used for extraction of $^{126}$Te cross-section (see section~\ref{online} for details)\\
b~:~extrapolated value used for extraction of $^{126}$Te cross-section
\end{table*}
The measured cross sections for residues from incomplete fusion; $\textit{viz.}$, $^{126,127}$Te and $^{124,125,126}$Sb along with one neutron stripping ($^{125}$Sn) and pickup ($^{123}$Sn) products are listed in Table~\ref{tab:table3} and plotted in Fig.~\ref{7Li_ICFdata}. The total $\textit{t}$-capture and total $\alpha$-capture cross sections are obtained from adding the individual residue cross-sections. The total $\textit{t}$-capture is found to be much larger than $\alpha$-capture at all the measured energies. Intuitively, we expect this behavior as triton while approaching the target sees lower Coulomb barrier compared to $\alpha$ particle. Hence the cross section for $\textit{t}$-capture is expected to be more compared to those of $\alpha$-capture. It is to be noted that deuteron and proton stripping from $^7$Li projectile would give the same ERs as those following $\textit{t}$-capture process and subsequent few neutron evaporation. Hence, from experiments it is difficult to separate these three processes.

In order to investigate the behaviour of observed residue cross-sections from $\textit{t}$-capture and $\alpha$-capture, the statistical model calculations were performed using PACE \cite{Gav80} code with modified prescription for level density \cite{Mahata15}. The spectrum of the surviving $\alpha$-particles, after capture of the complementary fragment (triton), represents the cross section for breakup-fusion as a function of the kinetic energy of the $\alpha$-particles. As seen from the literature \cite{Shrivastava13,Pfe73,Sig03,Santra12,Pradhan13} for $^{6,7}$Li induced reactions on various targets, the $\alpha$, deuteron and triton energy spectra have width, $\sigma$ $\sim$ 4 MeV centered around the 4/7 (for $\alpha$) and 3/7 (for triton) of beam energies in case of $^7$Li. Assuming Gaussian distribution, the whole $\alpha$ spectra (or excitation energy spectra of the intermediate nucleus formed after ICF) was divided into four bins of width 4 MeV each as in Ref.~\cite{Shrivastava13} with central two bins having 34\% weight and the outer two bins having 16\% weight. For each $^7$Li energy, the statistical model calculation was carried out for these four excitation energy bins and weighted sum was taken as the predicted cross section. The calculated values of absolute cross sections for the residues, $^{124,125,126}$Sb, are plotted in Fig.~\ref{7Li_ICFdata}(a) showing reasonably good agreement with the data. Following the same procedure, cross sections for residues arising from the capture of $\alpha$-particles were calculated from PACE with weight from the corresponding triton spectra. The results obtained are shown for $^{126,127}$Te residues in Fig.~\ref{7Li_ICFdata}(b) showing a similar agreement. The calculated cross section for $^{125}$Te is also shown in Fig.~\ref{7Li_ICFdata}(b). These calculations suggest that these residues are populated via fragment capture or transfer followed by evaporation, not through any other one step direct process.

\subsection{Simultaneous description of CF, ICF and TF cross sections}
\begin{figure}
\includegraphics[width=0.5\textwidth,trim=2.3cm 0.07cm 3.4cm 0.1cm, clip=true]{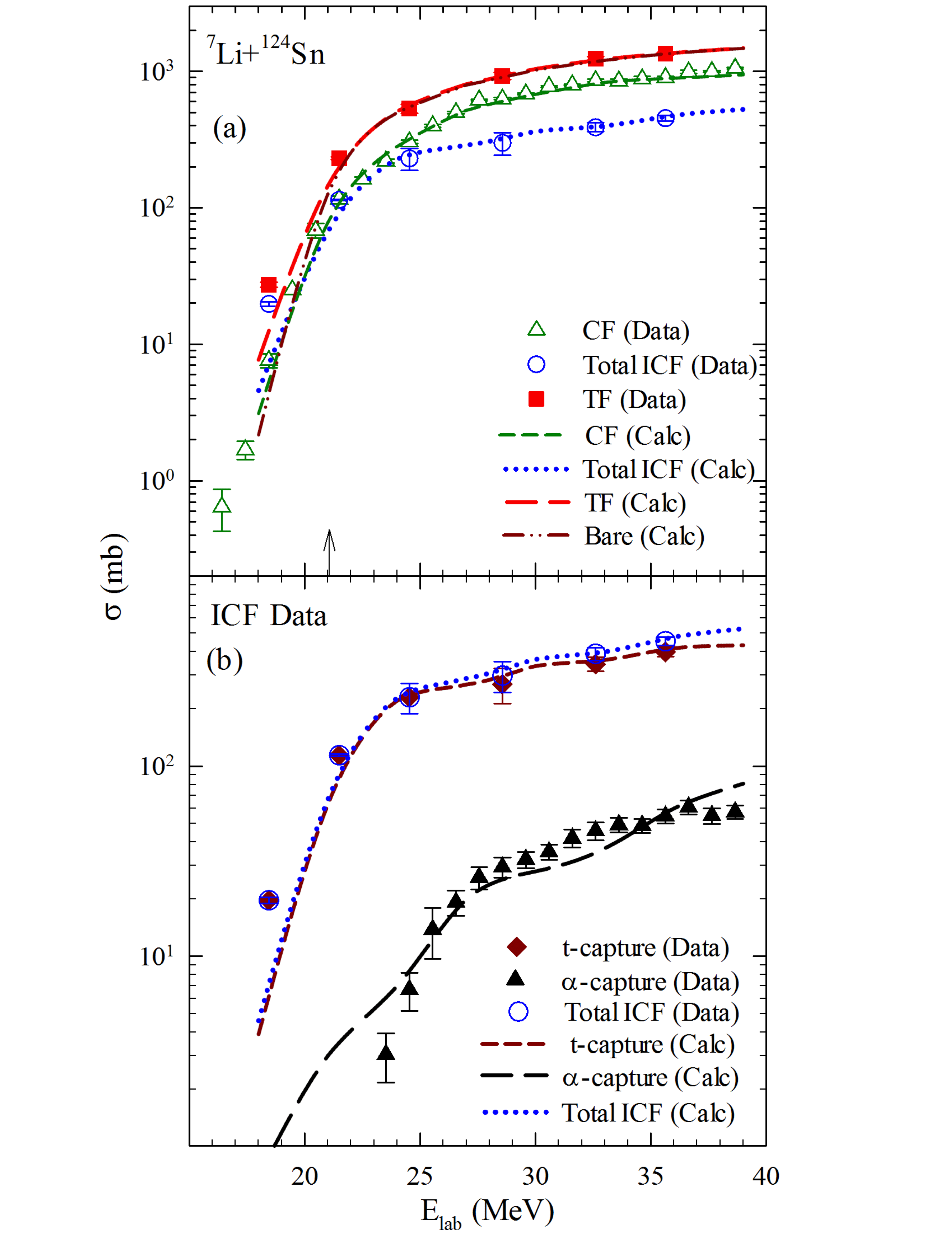}
\caption{\label{7Li_CFICFTF} (a) The data of CF, ICF and TF cross sections for $^{7}$Li+$^{124}$Sn reaction are compared with the coupled channel calculations. The arrow indicate the position of Coulomb barrier. (b) Comparison of individual ICF contributions from $\alpha$-capture, $\textit{t}$-capture along with Total ICF with the calculations. (see text for details).}
\end{figure}
\begin{figure}
\includegraphics[width=0.5\textwidth,trim=5.3cm 13.5cm 2.5cm 7.1cm, clip=true]{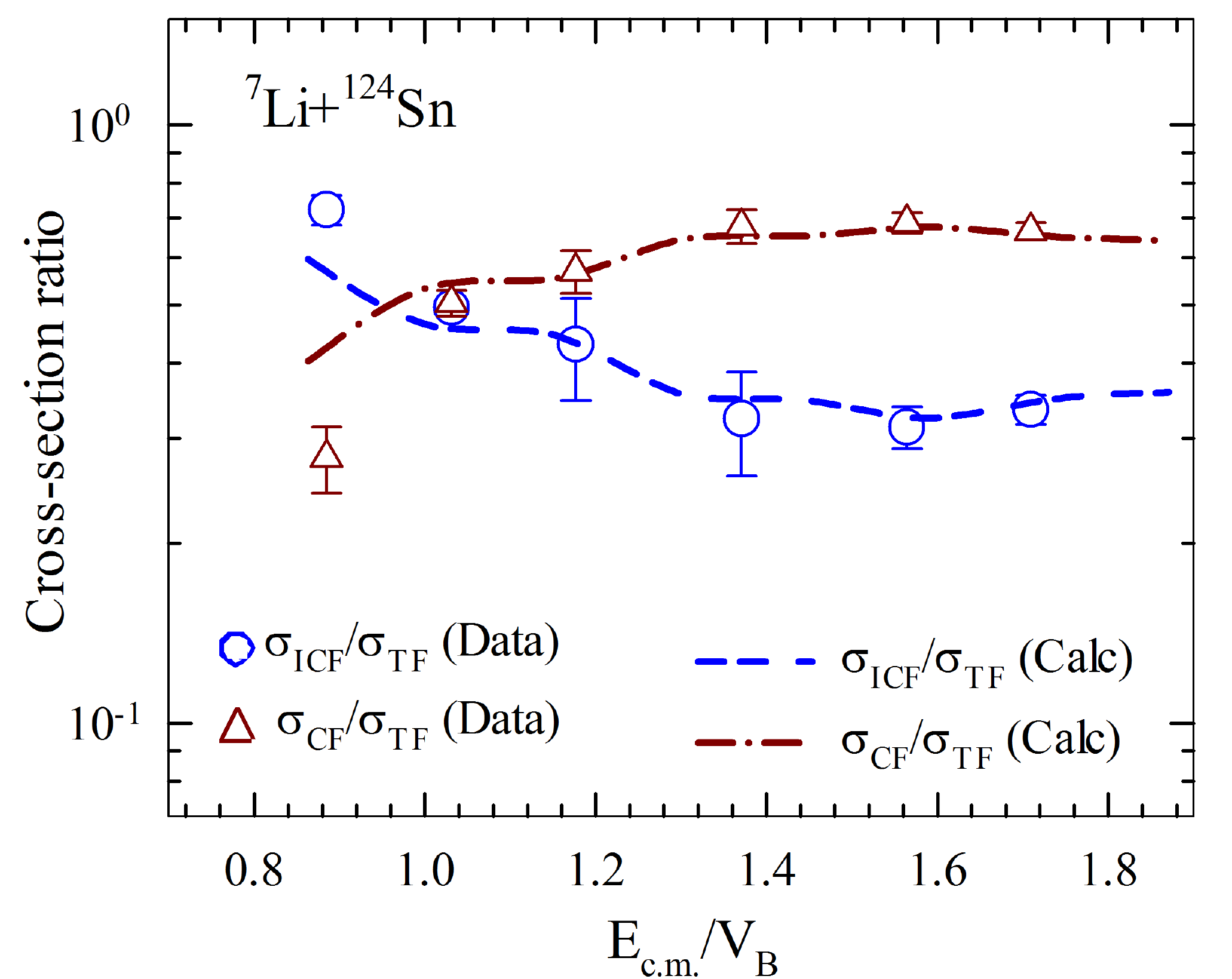}
\caption{\label{ratiofig} The ratio of cross sections, $\sigma_{\textrm{ICF}}/\sigma_{\textrm{TF}}$ and $\sigma_{\textrm{CF}}/\sigma_{\textrm{TF}}$ derived from the calculations as a function of E$_{\textrm{c.m.}}$/V$_{\textrm{B}}$ for $^{7}$Li+$^{124}$Sn reaction is shown by dashed and dashed-dot lines respectively. The symbols are showing the experimental data.}
\end{figure}
There have been some recent theoretical works, where separation of the ICF and CF components have been achieved using the calculations based on Continuum Discretized Coupled Channel (CDCC) formalism.  In Ref.~\cite{Diaz02}, CF and ICF cross sections are separated as the absorption from the projectile bound channels and the projectile breakup states respectively, where the absorption is calculated using a short range imaginary bare potential in the centre of mass motion. In another approach, two imaginary potentials are employed for interaction between the breakup fragments and target and the TF is defined as the cumulative absorption due to these potentials \cite{Diaz03,Jha14}. In the work of Hashimoto $\it{et~al.}$~\cite{Has09}, the CF is considered to arise when both the breakup fragments are in the range of  imaginary potentials whereas, the ICF arises when only one of the fragments is in the range of the respective imaginary potentials. They use the complete CDCC wavefunction with two imaginary potentials and utilize it for calculating the CF and ICF corresponding to absorptions in different regions. In this process, they use a radius  parameter  to divide the respective absorption regions and the CF and  ICF cross sections were calculated for the (d,p) reactions. In the work of Parkar $\it{et~al.}$~\cite{Parkar16}, the TF  and ICF cross sections were evaluated by modifying the absorption in an approximate way by selecting different set of short range imaginary potentials. A  sophisticated calculation method developed by Lei and Moro \cite{Lei15}, where they explicitly calculate the non-elastic breakup as the absorption of a given fragment when the other fragment survives by employing the proper outgoing boundary conditions.

Here we have followed the calculation method adopted in our earlier work \cite{Parkar16} where the detailed coupled channels calculations were performed using CDCC method using the code {\sc FRESCO} 2.9 \cite{Thomp88} for the simultaneous description of complete, incomplete and total fusion data for $^{6,7}$Li+$^{209}$Bi and $^{6,7}$Li+$^{198}$Pt reactions. Since in the present work for $^{7}$Li+$^{124}$Sn reaction, a complete set of data of CF, ICF and TF is available over a wide energy range, similar calculations are performed. The details of calculation method were already described in the earlier work \cite{Parkar16} and only the short summary regarding this work is presented here.

The binding potential for $\alpha$-t in $^7$Li was taken from Ref.~\cite{Buck88}, while the real part of required fragment-target potentials (V$_{\alpha-T}$ and V$_{t-T})$ in cluster folding model were taken from S$\tilde{a}$o Paulo potential \cite{Sau-Paulo}. In the calculations presented here, the fusion cross sections were first calculated by including the short-range imaginary (W$_{SR}$) volume type potentials in the coordinates of both projectile fragments relative to the target, as in Ref.\ \cite{Diaz03}. The short range imaginary potential for $\alpha$-T and t-T are: W$_0$ = 25 (25) MeV, r$_w$ = 0.60 (0.79) fm, a$_w$ = 0.4 (0.4) fm. Three set of  CDCC calculations with the breakup couplings were performed with three choices of optical potentials, where W$_{SR}$ was used for (i) both the projectile fragments relative to the target (Pot. A), (ii) the $\alpha$-T part only (Pot. B), and (iii) the t-T part only (Pot. C). In addition, an imaginary volume type potential with parameters W=25 MeV, r$_w$=1.00 fm and a$_w$=0.4 fm, without any real part was also present in the center of mass of the whole projectile for the projectile-target radial motion. The imaginary potential ensures that the total flux decreases by the absorption when the core and the valence cluster are in the range of the potential of target nucleus. Using the combination of the absorption cross sections with three potentials, the cross sections for (i) Total fusion ($\sigma_{\textrm{TF}}$), (ii) $\sigma_{\textrm{CF}}$+$\sigma_{\alpha}$, and (iii) $\sigma_{\textrm{\textrm{CF}}}$+$\sigma_{\textrm{t}}$ were calculated. These are further utilized  to estimate the  $\sigma_{\alpha\textrm{-capture}}$, $\sigma_{\textrm{t-capture}}$ and $\sigma_{\textrm{CF}}$ explicitly. The parameters of the short range imaginary potential in the range of r$_{w}$ = 0.6 to 1.0 fm and a$_{w}$ = 0.1 to 0.4 fm are found to be less sensitive for the calculation of $\sigma_{TF}$. However, in the calculation of ICF, the radius parameter of imaginary part is optimized with the higher energy ICF data.

In Fig.\ \ref{7Li_CFICFTF}(a) results of the calculations for the  TF, CF and ICF cross sections are shown with long dashed, short dashed and dotted lines, respectively along with the corresponding experimental data. The bare calculations (without breakup couplings) were also performed and the calculated fusion cross sections are denoted by dashed-dot-dot line. The Coulomb barrier position is marked by arrow in the figure. It is seen that at energies above the Coulomb barrier, the calculations which include the couplings and calculations that omit them have negligible difference but at energies below the barrier, the coupled TF cross sections are enhanced in comparison to bare TF cross sections. The calculated individual ICF cross sections, $\sigma_{\alpha\textrm{-capture}}$ and $\sigma_{\textrm{t-capture}}$, are shown in Figs.\ \ref{7Li_CFICFTF}(b) along with the measured data. In this figure, the long dashed, short dashed and dotted lines are the $\alpha$-capture, $\textit{t}$-capture and Total ICF calculations, respectively. The simultaneous description of CF, individual ICF and Total ICF was achieved from these coupled channels calculations. As can be seen from the Fig.~\ref{7Li_CFICFTF}(b), the $\textit{t}$-capture cross sections are much more dominant than $\alpha$-capture cross sections and almost equals to total ICF. Similar observation was also made in the recent work \cite{Parkar16} for $^7$Li+$^{209}$Bi and $^7$Li+$^{198}$Pt reactions. Here we point out that, experimentally the capture cross sections may include the  breakup and subsequent absorption in the target or the transfer followed by breakup and subsequent absorption in the target as explained in Refs.~\cite{Kalkal,Cook}. We have not considered the transfer followed by breakup and subsequent absorption explicitly as it is complicated process to model. Nevertheless, the breakup absorption as calculated here is supposed to model the ICF process in an effective way.

The ratio of cross sections, $\sigma_{\textrm{ICF}}/\sigma_{\textrm{TF}}$ and $\sigma_{\textrm{CF}}/\sigma_{\textrm{TF}}$ derived from the calculations as a function of E$_{\textrm{c.m.}}$/V$_{\textrm{B}}$ are shown by dashed and dash-dotted lines respectively in Fig.\ \ref{ratiofig}. The corresponding experimental data from the present measurement of $\sigma_{\textrm{ICF}}/\sigma_{\textrm{TF}}$ and $\sigma_{\textrm{CF}}/\sigma_{\textrm{TF}}$ are shown with hollow circles and hollow triangles respectively in Fig.\ \ref{ratiofig}. From the figure it is evident that (i) for the energies above the Coulomb barrier,  $\sigma_{\textrm{ICF}}/\sigma_{\textrm{TF}}$ and $\sigma_{\textrm{CF}}/\sigma_{\textrm{TF}}$ ratio remain approximately constant over the energy range and CF is dominant over the ICF  (ii) At the Coulomb barrier position, $\sigma_{\textrm{ICF}}/\sigma_{\textrm{TF}}$ is of similar magnitude as $\sigma_{\textrm{CF}}/\sigma_{\textrm{TF}}$ indicating the equal importance of CF and ICF and (iii) for energies below the barrier, the $\sigma_{\textrm{ICF}}/\sigma_{\textrm{TF}}$ is increasing while $\sigma_{\textrm{CF}}/\sigma_{\textrm{TF}}$ is decreasing showing the dominance of ICF over CF cross sections. The $\sigma_{\textrm{ICF}}/\sigma_{\textrm{TF}}$ ratio at above barrier energies gives the value of suppression factor in CF, which is found to be in agreement ($\sim$ 30 \%) with the literature data with $^7$Li projectiles from various measurements \cite{vvp10,Gas09,Wang14,Kundu16}. This value is direct experimental number for CF suppression factor and is matching with CCFULL calculations as shown in Section~\ref{sec:Coupled}. These results show that ICF is crucial for understanding the CF suppression factor.

\section{\label{sec:Sum} Summary}
The complete and incomplete fusion excitation function for $^7$Li+$^{124}$Sn reaction were measured in the energy range 0.80 $<$ V$_B$ $<$ 1.90 by online and offline $\gamma$-ray detection techniques. At above barrier energies, the measured complete fusion cross sections were found to be suppressed by a factor of 26 $\pm$ 4\% in comparison with the coupled channel calculations performed using the model adopted in CCFULL. This suppression factor is found to be in agreement with the literature data for the $^7$Li projectile on various targets and seem to suggest that the suppression factor does not vary appreciably at these energies for different target mass systems.  The measured $\textit{t}$-capture cross sections are significantly more than the $\alpha$-capture cross sections at all energies. Similar observations were also made on ICF data for $^7$Li+$^{209}$Bi and $^7$Li+$^{198}$Pt reactions in Ref.~\cite{Parkar16}. The statistical model calculations successfully explain the measured cross sections for the residues arising from the $\textit{t}$-capture and $\alpha$-capture underlining that the residues primarily originate from the two-step mechanism of breakup followed fusion-evaporation. The measured ICF cross sections taken as sum of $\textit{t}$-capture and $\alpha$-capture cross sections are  found to be commensurate with the suppression observed in the CF data. Further, simultaneous measurements of CF and ICF preferably in different target mass regions are required to understand these aspects.

We have also performed the CDCC based coupled channel calculations, which includes the coupling of breakup continuum of $^7$Li nucleus explicitly using the cluster folding potentials in the real part along with the short range imaginary potentials to calculate the CF, ICF and TF cross sections. The simultaneous explanation of the experimental data for the CF, ICF and TF cross sections over the entire energy range was obtained. The calculated TF cross-sections from uncoupled and coupled were found to match at energies above the barrier, while below barrier uncoupled TF is lower than the coupled one. The calculated and experimental ICF fraction, which is the ratio of ICF and TF cross sections is found to be constant at energies above the barrier and it increases at energies below the barrier showing the enhanced importance of ICF contribution in TF at below barrier energies. Further it will be of interest to describe this complete set of data using more sophisticated theories.

\begin{acknowledgments}
The authors would like to thank the Pelletron crew, Mumbai for the smooth operation of the accelerator during the experiment. We also thank Prof. J. Lubian for giving us the code to calculate S$\tilde{a}$o Paulo potentials. One of the authors (V.V.P.) acknowledges the financial support through the INSPIRE Faculty Program, from the Department of Science and Technology, Government of India, in carrying out these investigations.
\end{acknowledgments}

\end{document}